\documentclass[]{spie}  %>>> use for US letter paper
\usepackage{amsmath,amsfonts,amssymb}
\usepackage{graphicx}
\usepackage[hang,labelfont=it]{caption}
\usepackage{subcaption}
\usepackage[colorlinks=true, allcolors=black]{hyperref}

\title{Laboratory characterization of FIRSTv2 photonic chip for the study of substellar companions}

\author[a]{K. Barjot}
\author[a]{E. Huby}
\author[b,c]{S. Vievard}
\author[d]{N. Cvetojevic}
\author[a]{S. Lacour}
\author[e]{G. Martin}
\author[b,a]{V. Deo}
\author[a]{V. Lapeyrere}
\author[a]{D. Rouan}
\author[b,c,f]{O. Guyon}
\author[b]{J. Lozi}
\author[g]{N. Jovanovic}
\author[h]{C. Cassagenettes}
\author[a]{G. Perrin}
\author[i,a]{F. Marchis}
\author[e,j]{G. Duchêne}
\author[c,b]{T. Kotani}
\affil[a]{LESIA, Observatoire de Paris, Université PSL, CNRS, Sorbonne Université, Université de Paris, 5 place Jules Janssen, 92195 Meudon, France}
\affil[b]{National Astronomical Observatory of Japan, Subaru Telescope, 650 North Aohoku Place, Hilo, HI 96720, U.S.A.}
\affil[c]{Astrobiology Center of NINS, 2-21-1, Osawa, Mitaka, Tokyo, 181-8588, Japan}
\affil[d]{Universit\'e C\^ote d'Azur, Observatoire de la C\^ote d'Azur, CNRS, Laboratoire Lagrange, France}
\affil[e]{Universit\'e Grenoble Alpes / CNRS, Institut de Plan\'etologie et d’Astrophysique de Grenoble, 38000 Grenoble, France}
\affil[f]{College of Optical Sciences, University of Arizona, Tucson, AZ 85721, U.S.A.}
\affil[g]{California Institute of Technology, 1200 E California Blvd, Pasadena, CA 91125, U.S.A.}
\affil[h]{Teem Photonics, F-38240, Meylan, France}
\affil[i]{Carl Sagan Center at the SETI Institute, 189 Bernardo Av., Mountain View, CA 94043, USA}
\affil[j]{Astronomy Department, University of California at Berkeley, Berkeley, CA 94720, USA}

\authorinfo{Send correspondence to K. Barjot, E-mail: kevin.barjot@obspm.fr}

\begin{document} 
\maketitle

\begin{abstract}
 FIRST (Fibered Imager foR a Single Telescope instrument) is a post-AO instrument that enables high contrast imaging and spectroscopy at spatial scales below the diffraction limit. FIRST achieves sensitivity and accuracy by a unique combination of sparse aperture masking, spatial filtering by single-mode fibers and cross-dispersion in the visible. The telescope pupil is divided into sub-pupils by an array of microlenses, coupling the light into single-mode fibers. The output of the fibers are rearranged in a non redundant configuration, allowing the measurement of the complex visibility for every baseline over the 600-900 nm spectral range. A first version of this instrument is currently integrated to the Subaru Extreme AO bench (SCExAO). This paper focuses on the on-going instrument upgrades and testings, which aim at increasing the instrument’s stability and sensitivity, thus improving the dynamic range. FIRSTv2’s interferometric scheme is based on a photonic chip beam combiner. We report on the laboratory characterization of two different types of 5-input beam combiner with enhanced throughput. The interferometric recombination of each pair of sub-pupils is encoded on a single output. Thus, to sample the fringes we implemented a temporal phase modulation by pistoning the segmented mirrors of a Micro-ElectroMechanical System (MEMS). By coupling high angular resolution and spectral resolution in the visible, FIRST offers unique capabilities in the context of the detection and spectral characterization of close companions, especially on 30m-class telescopes.
\end{abstract}
% Short: \\
%   FIRST (Fibered Imager foR a Single Telescope) enables high contrast imaging and spectroscopy at spatial scales below the diffraction limit. FIRST achieves sensitivity and accuracy by a unique combination of sparse aperture masking, spatial filtering by single-mode fibers and cross-dispersion in the visible. The output of the fibers are re-arranged in a non redundant configuration, allowing the measurement of the complex visibility for every baseline over the 600-900nm spectral range. This presentation will focus on the on-going instrument upgrades, which aim at increasing the instrument’s stability and sensitivity, thus improving the dynamic range. We will present laboratory results obtained with two different types of 5-input photonic beam recombiner. Coupling high angular resolution and spectral resolution in the visible, FIRST offers unique capabilities in the context of the spectral characterization of close companions, thus can be extrapolated on a similar approach on a 30m-class telescope, which would provide unique scientific opportunities for companion detection and characterization.


%%%%%%%%%%%%%%%%%%%%%%%%%%%%%%%%%%%%%%%%%%%%%%%%%%%%%%%%%%%%%%%%
\section{Introduction}

%->>> Masquage de pupille, historique :---> imagerie a haute resolution angulaire (ex: Wolf rayet, Tuthill et al. 2000 \cite{tuthill2000aperture})

%->>> pour augmenter le contrast, et la transmission, concept avec fibre optique monomodes (Perrin et al. 2006 \cite{perrin2006high}, Lacour et al. 2006 \cite{lacour2007high})

%->>> premiere lumiere de l'instrument sur le Lick (E. Huby et al. 2012 \cite{huby2012first}): premier tests ciel, demonstration que l'on arrive a lambda/2D dans le visible

%->> Exoplanetes, context, GRAVITY peut detecter des exoplanetes (Gravity collaboration et al. 2019 sur HR8799e) grace a l'optique integre (IO)

%->>> new instrument for SUBARU, FIRST v2, base sur IO.

High contrast imaging at high angular resolution is crucial for the imaging and the spectroscopic study of faint stellar companions such as exoplanets. As the angular resolution is inversely proportional to the diameter of the telescope, larger telescopes are needed to improve it. Thus the interferometric combination of several telescopes (e.g. the four 8-meter VLT) is currently used to increase the resolution power in images. It is a successful technique but demands the use of several telescopes and the recombination of their light. As an alternative, the pupil masking technique\cite{tuthill2000aperture} proposes to interfere sub-divisions of one telescope pupil (via a mask with holes). It simplifies the scheme while improving the angular resolution up to twice the diffraction limit of the telescope. 
% on peut laisser comme ça, mais à mon avis on devrait plutot comparer le masquage aux performances d'une OA par exemple. L'interférométrie longue base atteint des résolutions angulaires qui sont vraiment plus fines que ce qu'on peut atteindre avec le masquage avec un seul télescope, du coup ce n'est pas tout à fait comparable. Pareil pour la sensibilité, surtout que tu cites la recombinaison des 4 UT, alors qu'en masquage on se retrouve à recombiner des sous pupilles de 1m de diamètre...

However, the sub-apertures have to be located non redundantly to minimize the degradation of the optical transfer function (OTF) of the telescope. As a consequence, it limits the dynamic range because only a small fraction of the pupil can be used. To overcome this issue and improve the image contrast, the pupil remapping concept has been proposed \cite{perrin2006high, lacour2007high}, where the entire surface of the pupil is divided into sub-apertures that are rearranged into a non-redundant pattern for the recombination. Single-mode optical fibers are used to perform this rearrangement and to filter the wavefront from aberrations.

The fibered imager for a single telescope (FIRST \cite{kotani2008first, huby2012first, vievard2020capabilities, Vievard2020FIRST}) is an instrument installed on the Subaru Coronagraphic Extreme Adaptive Optics (SCExAO \cite{2015PASP..127..890J}) at the Subaru Telescope, aiming at high angular resolution and high contrast imaging using pupil remapping with single-mode optical fibers. Prior to this, the instrument successfully validated these concepts with its first on-sky results \cite{huby2013first} at the Lick Observatory, demonstrating that images can be recovered at an angular resolution lower than the telescope diffraction limit at visible wavelengths.

In this paper, we present the upgrade and laboratory characterization of the second version of the instrument, FIRSTv2, now including a photonic chip, in order to improve the stability and sensitivity. Integrated optics (IO) is now a key technology to perform the interferometric combination in a small volume. As an example, the GRAVITY instrument operating at the very large telescope interferometer (VLTI) is based on a photonic recombination \cite{perraut2018single} and allows the characterization of exoplanets \cite{lacour2019first}.

%%%%%%%%%%%%%%%%%%%%%%%%%%%%%%%%%%%%%%%%%%%%%%%%%%%%%%%%%%%%%%%%
\section{Instrument description}

%%%%%%%%%%%%%%%%%%%%%%%%%%%%%%%%
\subsection{FIRSTv2 testbed}

FIRSTv2 is a laboratory testbed and is the upgrade of the FIRST instrument installed on the SCExAO bench at the Subaru Telescope since 2013. It aims at improving the contrast obtained on astrophysical data at high resolution.

\begin{figure}[!h]
    \centering
    \includegraphics[width=1.06\linewidth]{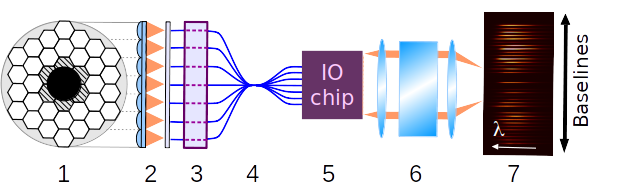}
    \caption{Schematic of the FIRSTv2 instrument. From left to right are shown: (1) the pupil sampling part, where the blue sub-pupils are the ones which light is injected into fibers, (2) the micro-lenses for the injection into the optical fibers, (3) the optical delay lines, (4) the single-mode fibers, (5) the photonic chip for the recombination, (6) the prism and (7) the final image on the camera.}
    \label{fig:FIRSTv2Scheme}
\end{figure}

Fig.~\ref{fig:FIRSTv2Scheme} depicts a schematic of the FIRSTv2 instrument. Denoted by (1) is the telescope pupil (with only a central obstruction here) represented on top of the 37-segment deformable mirror which sub-divides the pupil into sub-pupils. The blue sub-pupils are the ones that are used in the FIRSTv2 experiment. Since optical delay lines (ODL) have been included to control the optical path length differences (3), the configuration of these sub-pupils can actually be modified at will. Then (2) are the micro-lenses that inject the light of each sub-pupil into single-mode optical fibers (4) which filter the atmospheric wavefront aberrations. Only the differential piston remains. The photonic chip \cite{Guillermo2020BeamCombiner} (5) performs the recombination between the sub-pupils (see section~\ref{sec:Photonic}). Finally, the prism (6) cross-disperses the light before being focused onto the camera (7).

The FIRSTv2 upgrade consists in improving the contrast of the instrument by: 
\begin{itemize}
    \item[$-$] performing the interferometric recombination in a photonic chip instead of at the focal plane (on the camera),
    \item[$-$] improving the search for the fringes with the use of ODLs to accurately equalize the optical path length of all optical fibers,
    \item[$-$] encoding each interference pattern for a given baseline on few pixels instead of hundreds (as shown in Fig.~\ref{fig:FIRSTv2Scheme} (7)) which increases the sensitivity of the instrument.
\end{itemize}

%%%%%%%%%%%%%%%%%%%%%%%%%%%%%%%%
\subsection{Interferometric part}
\label{sec:Photonic}

\begin{figure}[!h]
    \begin{subfigure}{.5\textwidth}
        \includegraphics[width=\linewidth]{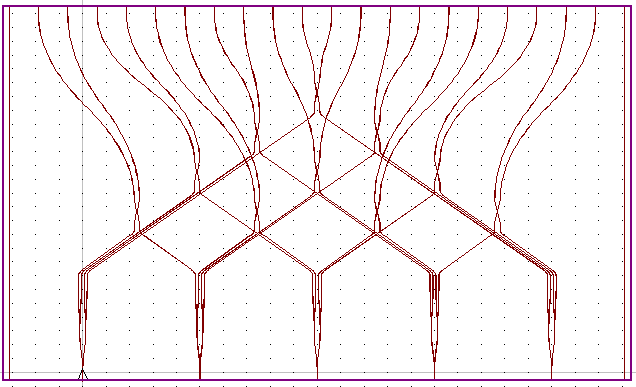}
    \end{subfigure}%
    \begin{subfigure}{.5\textwidth}
        \includegraphics[width=\linewidth]{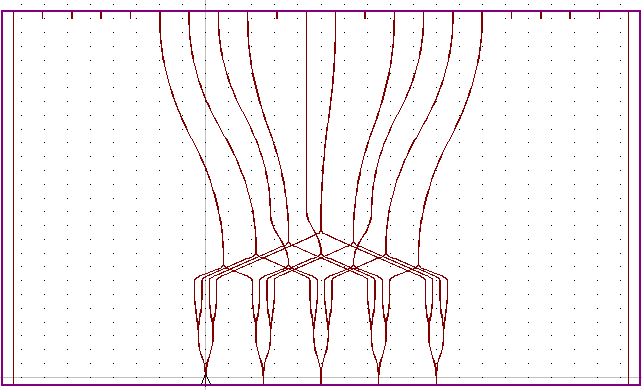}
    \end{subfigure}
    \caption{Schematics of the two photonic chips we characterized for FIRSTv2. The inputs are located at the bottom and the outputs at the top. Left is a \textit{X coupler} type and right is a \textit{Y coupler} type.}
    \label{fig:PhotonicScheme}
\end{figure}

The photonic chip is manufactured by Teem Photonics\footnote{https://www.teemphotonics.com}. It consists in a block of glass in which optical wave guides are engraved by photolithography. For this experiment, they have been optimized for $650 \, nm$. The chips have a number $n_{inputs}$ of inputs corresponding to the number of sub-pupils that interfere with each other. As shown in Fig.~\ref{fig:PhotonicScheme}, the waveguide corresponding to each input is divided into $n_{inputs} - 1$ waveguides, such that every input beam interferes with all the others. We have tested two types of combiner chip: the \textit{X-coupler} type (left schematic of Fig.~\ref{fig:PhotonicScheme}), leading to two outputs per recombined baseline, i.e. $n_{outputs} = n_{inputs} (n_{inputs} - 1) = 5 \times 4 = 20$, and the \textit{Y-coupler} type (right schematic of Fig.~\ref{fig:PhotonicScheme}), leading to one output per recombined baseline, i.e. $n_{outputs} = n_{inputs} (n_{inputs} - 1) / 2 = (5 \times 4) / 2 = 10$.
%Hence the outputs of the chip are all the possible combinations between the inputs, which are called baselines and can be numbered as $n_{outputs} = n_{inputs} (n_{inputs} - 1) = 5 \times 4 = 20$ for the \textit{X coupler} type (left schematic of Fig.~\ref{fig:PhotonicScheme}) or $n_{outputs} = n_{inputs} (n_{inputs} - 1) / 2 = (5 \times 4) / 2 = 10$ for the \textit{Y coupler} type (right schematic of Fig.~\ref{fig:PhotonicScheme}).

%%%%%%%%%%%%%%%%%%%%%%%%%%%%%%%%%%%%%%%%%%%%%%%%%%%%%%%%%%%%%%%%
\section{Photonic chip characterization}

%%%%%%%%%%%%%%%%%%%%%%%%%%%%%%%%
%\subsection{Measurement method}
%\label{sec:MeasMethod}

To characterize the two photonic chips designed and manufactured for FIRSTv2, we use a broadband Halogen light source \textit{Ocean Optics, HL-2000-FHSA-HP}\footnote{https://www.oceaninsight.com/globalassets/catalog-blocks-and-images/manual--instruction-re-branded/light-sources/mnl-1013-hl-2000-fhsa-users-manual-rev-a.pdf}. By putting the source light through an optical fiber focused on the camera, its spectrum is used to normalized the spectra obtained by injecting the light into the chips, in order to estimate their throughput as a function of wavelength. Three features of the chips were assessed: (1) the flux cross-talk between the outputs, (2) the throughput and (3) the contrast performances.

%%%%%%%%%%%%%%%%%%%%%%%%%%%%%%%%
\subsection{Cross-talk measurement}

This measurement aims at quantifying the unwanted light leaking from one waveguide to others, in particular at the location of the couplers. In the case where light is injected into one input only, only 4 or 8 outputs are expected to show non-zero flux for the Y-coupler and X-coupler chip respectively. The unwanted flux measured in the other outputs corresponds to the cross-talk leakage.

\begin{figure}[!h]
    \begin{subfigure}{\textwidth}
        \centering
        \includegraphics[width=0.8\linewidth]{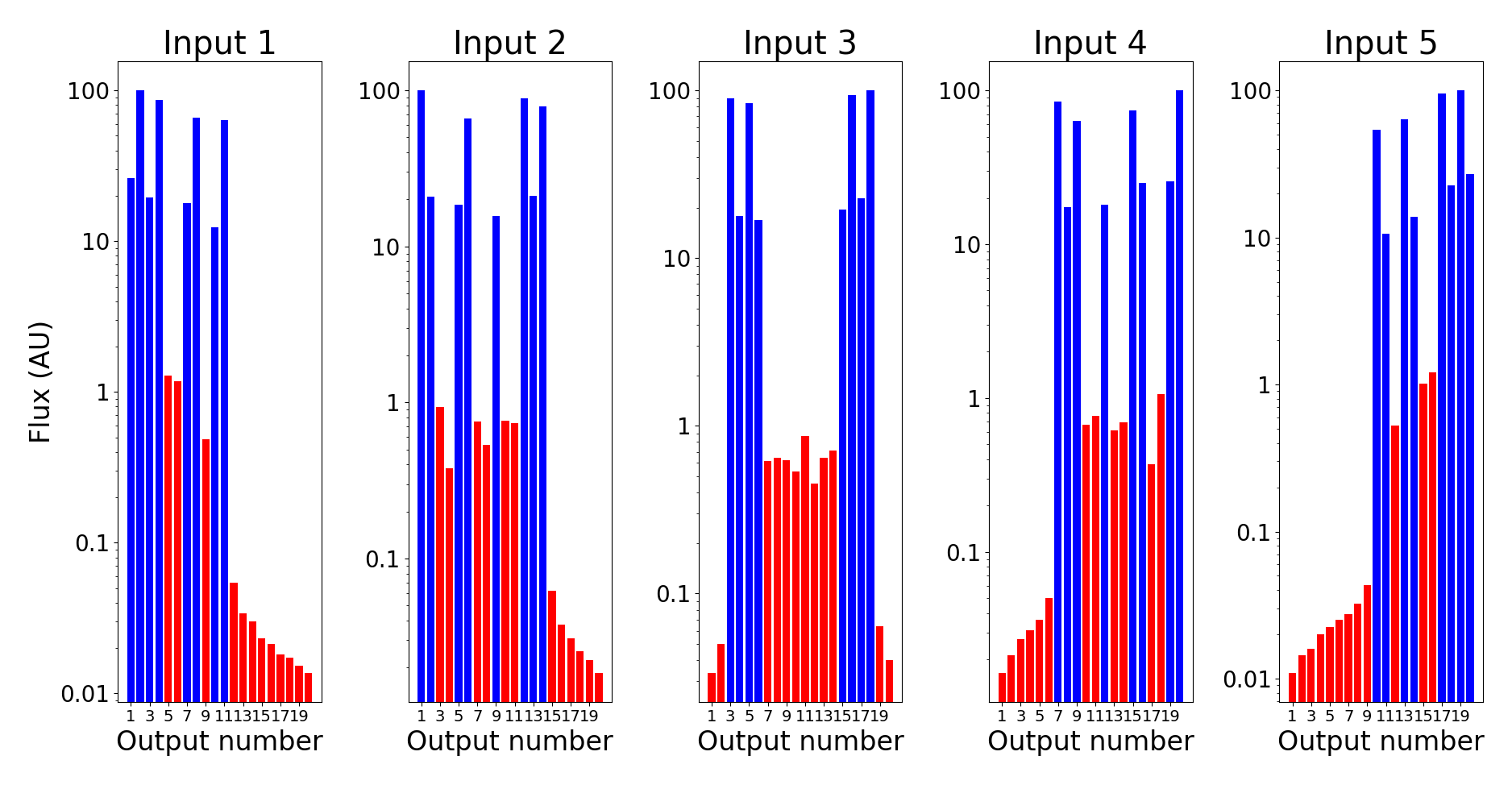}
    \end{subfigure}
    \begin{subfigure}{\textwidth}
        \centering
        \includegraphics[width=0.8\linewidth]{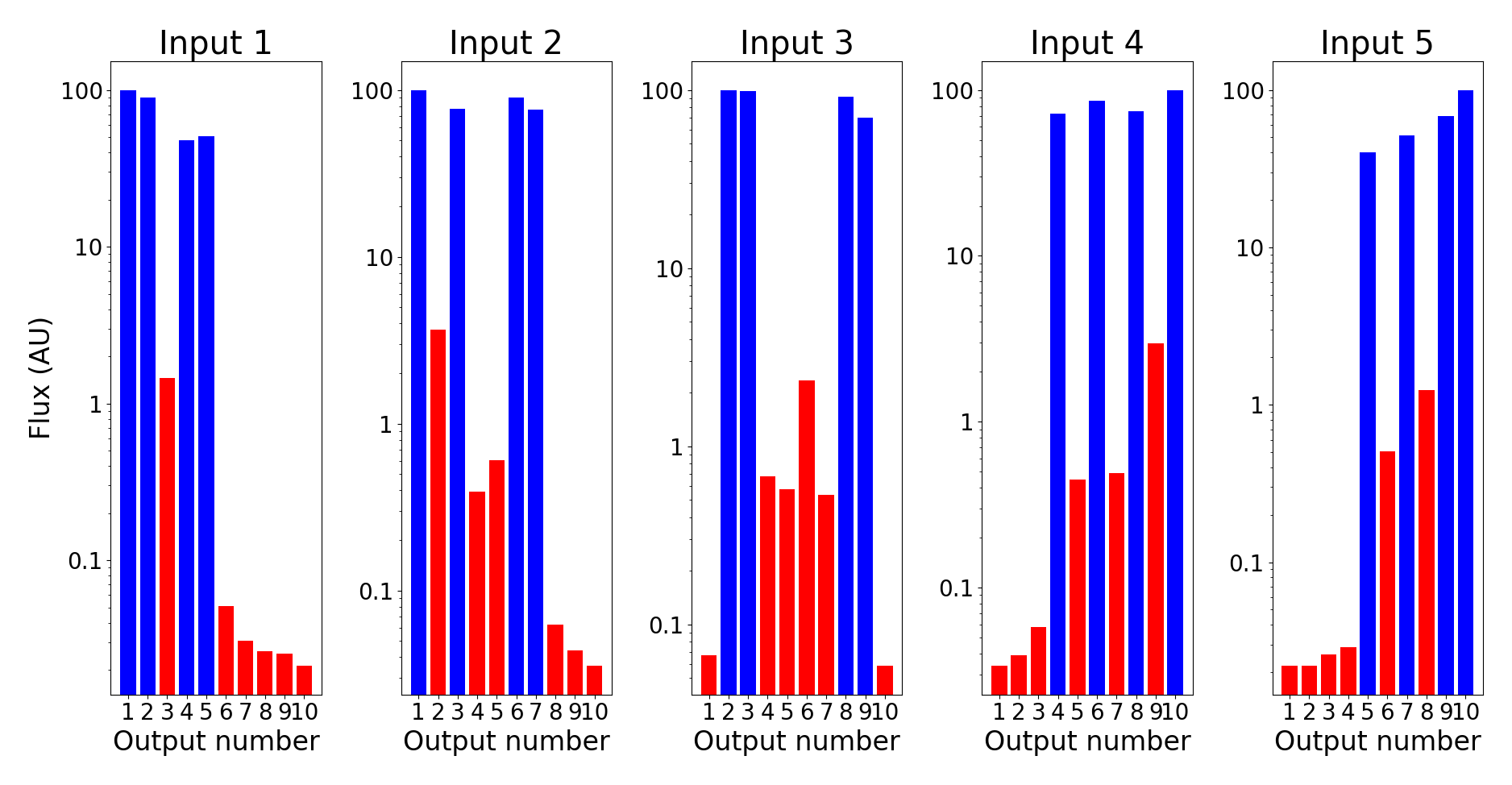}
    \end{subfigure}
    \caption{
    Cross-talk leakage characterization for the X-coupler (top) and the Y-coupler (bottom) chips. For each chip, there are five bar-plots, showing the flux measured in all outputs, while light is injected in one input only (the illuminated input corresponds to the number in the title). The blue bars represent the outputs where light is expected, while the red bars represent the other outputs expected to have no light.
    }
    \label{fig:CrosstalkCharac}
\end{figure}

Fig.~\ref{fig:CrosstalkCharac} presents the cross-talk characterization obtained for the X-coupler chip on top and for the Y-coupler chip at the bottom. For each chip, there are five bar-plots (titled from input 1 to input 5) showing the fluxes measured on all outputs while the \textit{Ocean Optics} source is injected in one input. The blue bars are the outputs where light is expected, while the red bars are the other outputs expected to show no light. There is at worst a $10 \%$ cross-talk for the X-coupler chip and a $20 \%$ cross-talk for the Y-coupler chip.

%%%%%%%%%%%%%%%%%%%%%%%%%%%%%%%%
\subsection{Throughput comparison}

\begin{figure}[!h]
    \centering
    \includegraphics[width=0.8\linewidth]{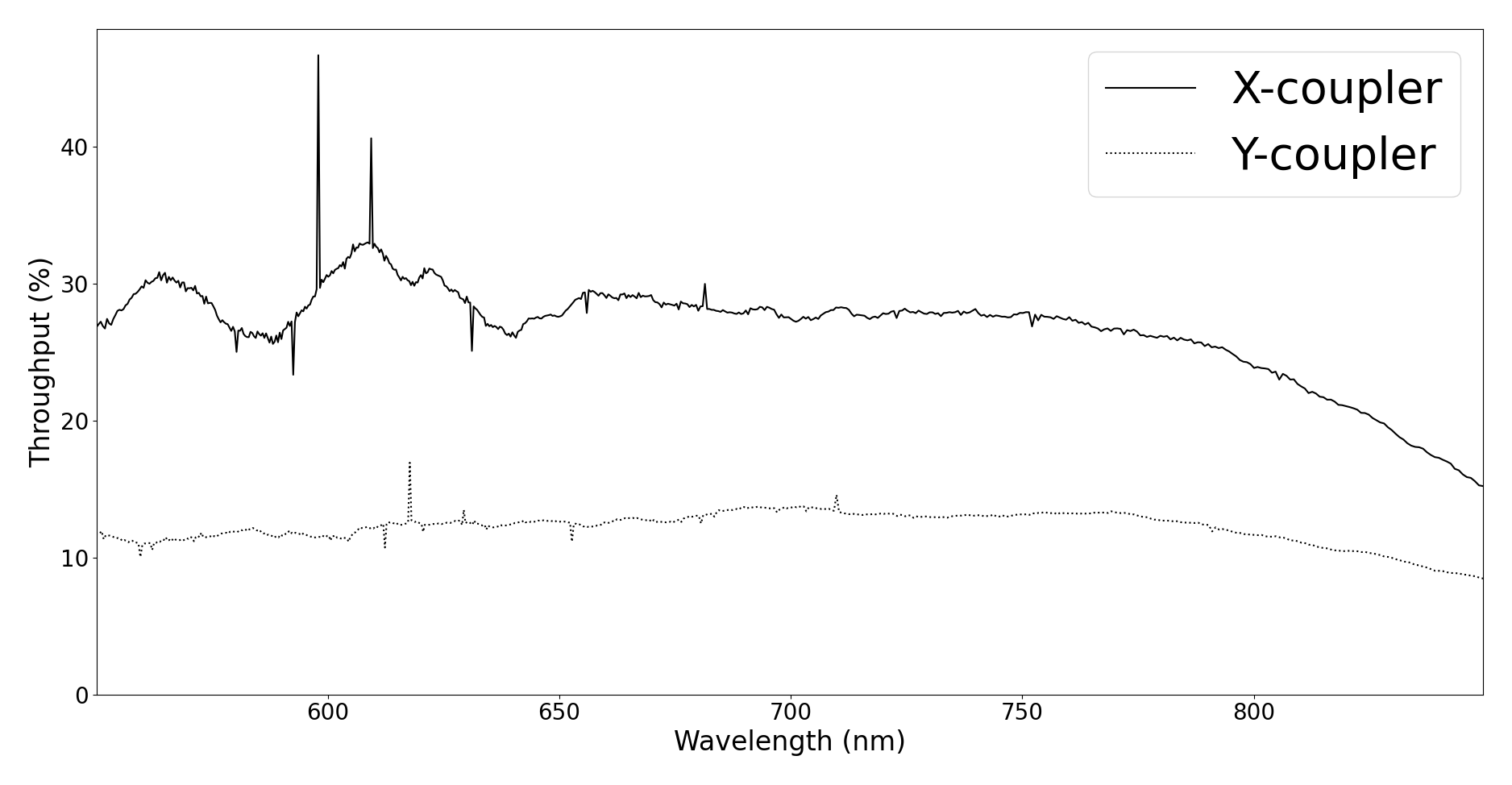}
    \caption{Measured throughput as a function of wavelength for the X- and Y-coupler chips in continuous and doted lines, respectively.}
    \label{fig:ThroughputCharac}
\end{figure}

The measured throughput are presented on Fig.~\ref{fig:ThroughputCharac} for the two types of couplers. Between $\sim 600 \, nm$ and $\sim 850 \, nm$ (working bandwidth of the chips) the throughput of the X-coupler represented by the continuous line is measured to $\sim 30 \, \%$ and the throughput of the Y-coupler represented by the doted line is measured to $\sim 13 \, \%$. This throughput levels are satisfying in comparison with the throughput of our first test chips reaching no more than $1 \, \%$ throughput.

%%%%%%%%%%%%%%%%%%%%%%%%%%%%%%%%
\subsection{Expected fringe contrast}

When illuminating two inputs with light intensities $I_1$ and $I_2$, their interferometric combination is measured on one of the outputs with an intensity $I$ given by:

\begin{equation}
\label{eq:InterferoIntensity}
I = I_1 + I_2 + 2 \sqrt{I_1 I_2} \cos{\left( \frac{2\pi \times OPD}{\lambda} \right)}
\end{equation}

where $OPD$ is the optical path length difference between the two beams and $\lambda$ the wavelength.

From equation~\ref{eq:InterferoIntensity}, the contrast of the fringes is obtained by the factor $2 \sqrt{I_1 I_2}/(I_1 +I_2)$. Thus we can use this definition to estimate the contrast performance expected for each baseline of the chips using the flux values of Fig.~\ref{fig:CrosstalkCharac}. These measurements were performed with only one input illuminated at a time. Hence, the expected contrast for baseline $i$ is assessed with the combination of two fluxes represented by two blue bar-plots. For instance, the contrast of the fifth baseline of the X-coupler chip is obtained from the combination of the flux measured on the fifth outputs of the bar-plot titled \textit{input 2} and of the bar-plot titled \textit{input 3}, normalised by the sum of the two intensities.

\begin{figure}[!h]
    \centering
    \includegraphics[width=0.8\linewidth]{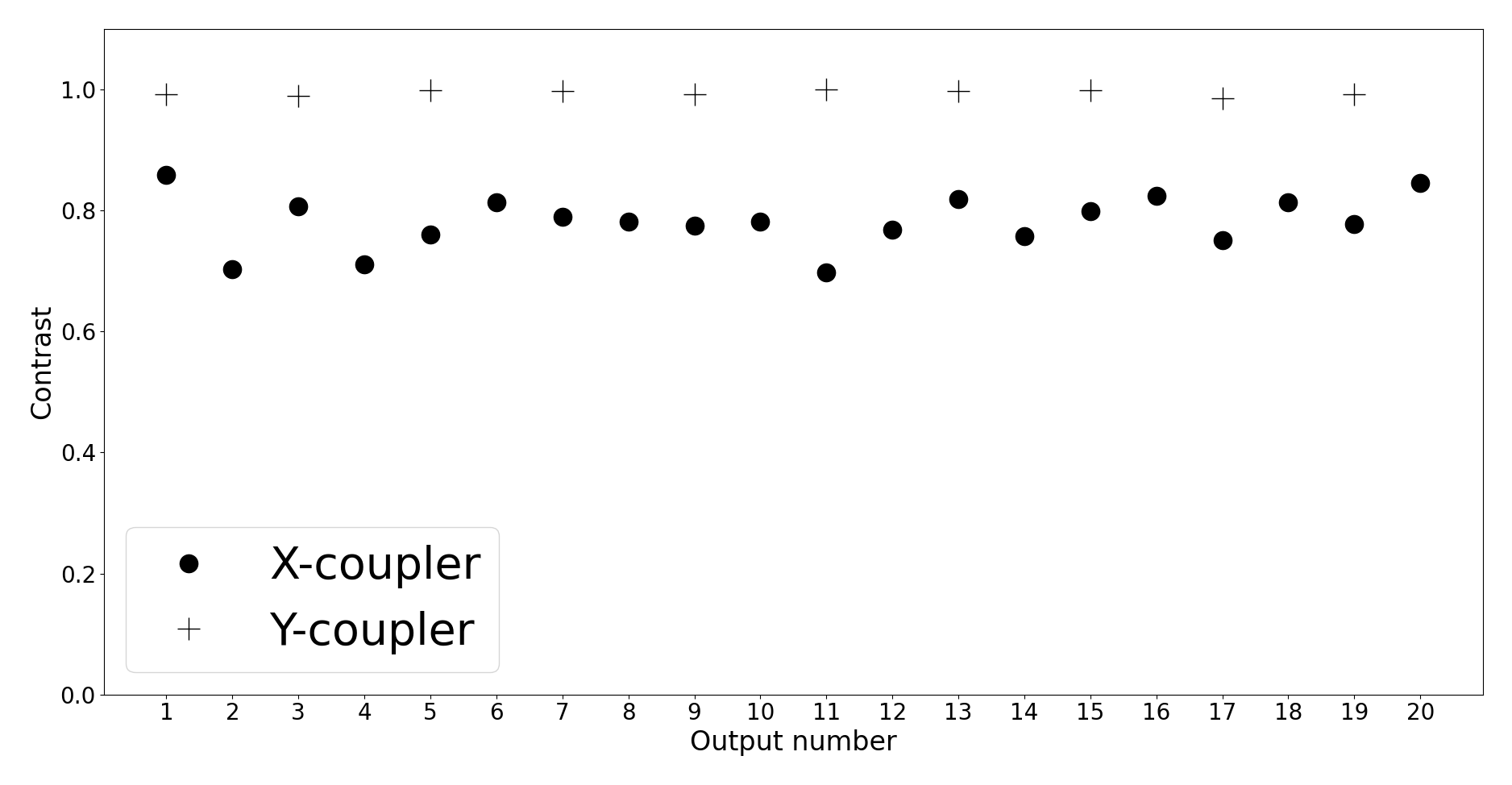}
    \caption{Interferometric contrasts obtained from the flux intensities of the Fig.~\ref{fig:CrosstalkCharac} as a function of the baselines for the X-coupler in continuous line and for the Y-coupler in doted line.}
    \label{fig:ContrastCharac}
\end{figure}

Fig.~\ref{fig:ContrastCharac} shows the expected fringe contrast estimated from the combination of every pair of output fluxes, shown in Fig.~\ref{fig:CrosstalkCharac}. Thus, the best contrast that can be obtained is measured to $2 \sqrt{I_1 I_2} = 2 \sqrt{6.0 \times 10^6 \times 1.8 \times 10^7} / (2.4 \times 10^7) \approx 0.866$ with the X-coupler and to $2 \sqrt{I_1 I_2} = 2 \sqrt{1.8 \times 10^7 \times 1.6 \times 10^7} / (3.3 \times 10^7) \approx 0.998$ with the Y-coupler.%, which is twice the first one.

It appears from this study that the X-coupler type has twice the throughput ($\sim 30 \, \%$) of the Y-coupler type ($\sim 13 \, \%$) but has lower contrast (contrast equals to $\sim 0.87$ at best for the first one and equals to $\sim 1$ for the second one). As the upgrade of FIRST aims at improving the dynamic range in the measurements, the Y-coupler chip is preferable to optimise the contrast performances.

%%%%%%%%%%%%%%%%%%%%%%%%%%%%%%%%%%%%%%%%%%%%%%%%%%%%%%%%%%%%%%%%
\section{Conclusion}

Based on the pupil remapping technique with photonic technology, FIRSTv2 is currently under lab characterization. In this paper we have shown a comparison between two chips performing the interferometric combination of pairs of beams by two different types of couplers: X-type and Y-type. Our study shows that the first one is twice better in transmission but has a lower performance in contrast than the second. As a consequence the work to come on FIRSTv2 will focus on the improvement of the Y-type photonic chip throughput.

%%%%%%%%%%%%%%%%%%%%%%%%%%%%%%%%%%%%%%%%%%%%%%%%%%%%%%%%%%%%%%%%
\bibliography{report} % bibliography data in report.bib
\bibliographystyle{spiebib} % makes bibtex use spiebib.bst

\end{document}